\documentclass[iop,apj,numberedappendix]{emulateapj}
\usepackage[latin1]{inputenc}
\usepackage{amssymb}
\usepackage{amsmath}
\usepackage{graphicx}
\usepackage{xspace}
\usepackage{natbib}
\usepackage{url}
\usepackage{paralist}

\slugcomment{ }

\shorttitle{Black hole spin and soft excess of Fairall~9}
\shortauthors{Lohfink et al.}

\begin{document}

\title{The black hole spin and soft X-ray excess of the luminous Seyfert galaxy Fairall~9}


\author{Anne M. Lohfink\altaffilmark{1,2}, Christopher S. Reynolds\altaffilmark{1,2}, Jon M. Miller\altaffilmark{3}, Laura W. Brenneman\altaffilmark{4}, Richard F. Mushotzky\altaffilmark{1,2}, Michael A. Nowak\altaffilmark{5}, Andrew C. Fabian\altaffilmark{6}} 

\altaffiltext{1}{Department of Astronomy, University of Maryland, College Park, MD 20742-2421, USA; alohfink@astro.umd.edu}
\altaffiltext{2}{Joint Space-Science Institute (JSI), College Park, MD 20742-2421, USA}
\altaffiltext{3}{Department of Astronomy, University of Michigan, 830 Dennison Bldg., 500 Church St., Ann Arbor, MI 48109-1042, USA}
\altaffiltext{4}{Harvard-Smithsonian Center for Astrophysics, 60 Garden Street, Cambridge, MA, USA}
\altaffiltext{5}{Massachusetts Institute of Technology, Kavli Institute for Astrophysics, Cambridge, MA 02139, USA} 
\altaffiltext{6}{Institute of Astronomy, University of Cambridge, Madingley Road, Cambridge CB3 0HA, UK}

\begin{abstract}
We present an analysis of all {\it XMM-Newton} and {\it Suzaku} X-ray spectra of the nearby luminous Seyfert galaxy Fairall~9.  Confirming previous analyses, we find robust evidence for a broad iron line associated with X-ray reflection from the innermost accretion disk.  By fitting a spectral model that includes a relativistically ionized reflection component, we examine the constraints on the inclination of the inner accretion disk and the black hole spin, and the complications introduced by the presence of a photoionized emission line system.  Employing multi-epoch fitting, we attempt to obtain robust and concordant measures of the accretion disk parameters.  We also clearly see a soft X-ray excess in Fairall~9. During certain epochs, the soft excess can be described with same disk reflection component that produces the iron line. However, there are epochs where an additional soft component is required. This can be attributed to either an additional highly-ionized, strongly blurred disk reflection component, or a new X-ray continuum component. 
\end{abstract}



\section{Black Hole Spin Constraints from AGN}\label{intro}

X-ray spectroscopy provides one of the best tools for studying the properties of astrophysical black holes. The hard X-ray tail in active galactic nuclei (AGN) and Galactic Black Hole Binaries (GBHBs) is produced by the inverse Compton scattering of soft thermal photons from the accretion disk by hot/energetic electrons in the disk corona.   Tremendous insight can be gained by studying the reflection/reprocessing of these coronal X-rays from the accretion disk \citep{fabian:89a,reynolds:03a,miller:07a}.  Identifying and characterizing the disk reflection features in the X-ray spectrum enables us to study the physical properties of the inner accretion disk as well as the spin of the black hole itself. This is possible because the reflection spectrum is affected by strong relativistic effects, specifically Doppler shift and gravitational redshift which give all emission line features a characteristic skewed profile.   These line profiles, especially that of the strong and isolated iron-K$\alpha$ fluorescent line, can reveal the location of the innermost stable circular orbit (ISCO) and hence the black hole spin.

Along with its mass, spin is the most fundamental parameter describing a black hole and has a significant influence on physics close to the black hole.  Importantly, in an AGN or GBHB, black hole spin is the only major source of energy in the system other than accretion and is often cited as the likely power source for relativistic jets \citep{blandford:77a}.   Moreover, the distribution of spin across a population of black holes provides a new way of probing the formation and growth of supermassive black holes \citep{volonteri:10a}.  However, in order to examine these astrophysical consequences of black hole spin, it is crucial that we have robust ways of measuring spin.

In the AGN realm, there are now a number of objects for which the supermassive black hole (SMBH) spin has been measured using X-ray reflection; for a summary, see Table~2 in \citet{brenneman:11a}.
Many of these previous studies have considered objects that show strong warm absorbers in their soft X-ray spectra. This has led some authors to suggest that the detection of relativistically broadened iron lines is an artifact of incorrectly modeling the absorption \citep{miller:08a}. While arguments exist against these absorption dominated models \citep{reynolds:09a}, it is clearly interesting to study relativistic disk reflection in objects without any obvious absorption. Therefore, in this paper, we study one of the brightest ``bare" Seyfert galaxies to show relativistic disk reflection but no warm absorber -- Fairall~9. 

Fairall~9 is a well-studied Seyfert 1 galaxy ($z=0.047$) that has never shown any imprints of a warm absorber in its X-ray spectrum \citep{reynolds:97a,gondoin:01a,schmoll:09a,emmanoulopoulos:11a}. The Galactic column in the direction of Fairall is also modest \citep[$3.15\times10^{20}\,\text{cm}^{-2}$;][]{kalberla:05a}, giving us an unobscured view of its soft X-ray continuum.  Fairall~9 shows limited X-ray variability on both long and short timescales \citep{markowitz:01a} with the exception of occasional sudden dips that last one or two weeks \citep{lohfink:12a}.  Studies with {\it ASCA} and early {\it XMM-Newton} observation  revealed a prominent narrow, neutral iron K$\alpha$ line and a possibly broad iron line \citep{reynolds:97a,gondoin:01a}.    An analysis of the 2007 \textit{Suzaku} observation by \citet{schmoll:09a} confirms the presence of a moderately weak broad iron line and estimates the dimensionless spin parameter of the black hole to be $a=0.6\pm 0.07$ under the assumption that the emissivity profile of the X-ray reflection is an unbroken power law with an emissivity index constrained to be less than 5.    More recently, \citet{emmanoulopoulos:11a} analyzed a moderately deep 2010 \textit{XMM-Newton} pointing (130\,ks), and found a weak spin constraint ($a=0.39^{+0.48}_{-0.30}$) that spans the range from slowly spinning to rather rapid prograde spin.   Worryingly, the disk inclination of $i=64^{+7}_{-9}$\,deg derived by \citet{emmanoulopoulos:11a} is strongly inconsistent with the value of $i=44\pm 1$\,deg measured by \cite{schmoll:09a}.   Most recently, \citet{patrick:11a} have examined a deep 2010 {\it Suzaku} pointing obtained as part of the {\it Suzaku AGN Spin Survey} Key Project (PI Reynolds).   They found the spin to be unconstrained when the spectrum is fitted by a blurred ionized disk spectrum, and determined yet a third value for the disk inclination, $i=33^{+3}_{-5}$\,deg.

In this {\it Paper}, we aim to bring clarity to the question of the black hole spin and disk inclination in Fairall~9 by considering all available \textit{XMM-Newton} \& \textit{Suzaku} pointings, including the long \textit{XMM-Newton} pointing of 2010 and the \textit{Suzaku} Key Project pointing of 2010.    To obtain robust measurements of the accretion disk parameters and black hole spin it is crucial to understand the continuum and its variation to the highest precision possible. We highlight the importance of multi-epoch fitting techniques in breaking a modeling degeneracy resulting from the presence of narrow ionized iron lines, as well as uncovering evidence for a soft excess above and beyond that resulting from soft X-ray reflection from the a uniformly ionized accretion disk.  Using this methodology, we derive concordant values for black hole spin and disk inclination.  

The outline of this paper is as follows. First, we describe the datasets used in this work and briefly discuss data reduction techniques (\S \ref{data}). After a preliminary investigation of spectral fits to the individual data epochs, we perform a multi-epoch analysis in order to produce robust and concordant constraints on on the spin of the black hole and the inclination of the disk (\S \ref{spectral}).   We end with a brief discussion of the lessons learned concerning methodologies for extracting black hole spin from X-ray spectra (\S \ref{dis}).  

\section{Observations \& Data Reduction}\label{data}

\begin{table*}[t]
\begin{center}
\caption{Overview of observations and exposures.\label{tab1}   Superscript$^{\rm a}$ denotes that XIS0/XIS1/XIS3 were used.   All {\it Suzaku} data were obtained after the failure of XIS2.}
\begin{tabular}{c c c c c}
\hline
\hline
Observatory & Instrument & Date & ObsID & Exposure  \\
&  &  &  & [ksec] \\ 
\tableline
\textit{Suzaku} (B) & XIS\tablenotemark{a} & 2010/05/19 & 705063010 & 191 \\
& PIN &  & & 162 \\
\textit{Suzaku} (A) & XIS\tablenotemark{a} & 2007/06/07 & 702043010 & 139  \\
& PIN &  & & 127 \\
\textit{XMM-Newton} (B) & EPIC-pn & 2009/12/09 & 0605800401 & 91 \\
\textit{XMM-Newton} (A) & EPIC-pn & 2000/07/05 & 0101040201 & 16 \\
 \tableline
\end{tabular}
\end{center}
\end{table*}

\subsection{Data Reduction}
\subsubsection{Suzaku}\label{suz_red}
 
A summary of the observations discussed in this paper is presented in Table~\ref{tab1}.   All \textit{Suzaku} data reduction was performed within Heasoft v6.11. The version 20110608 of the relevant \textit{Suzaku} CALDB files were used.  The \textit{Suzaku} data were reduced following the standard procedure described in the ``\textit{Suzaku} ABC Guide\footnote{http://heasarc.nasa.gov/docs/suzaku/analysis/abc/abc.html}". 

For the \textit{Suzaku} X-ray Imaging Spectrometer \citep[XIS;][]{koyama:07a}, all data were taken in the full window mode. The XIS data files were first reprocessed as recommended by the ABC Guide. For all operating XIS detectors (XIS0, XIS1, XIS3), the source extraction region was chosen to be a circle of radius of $\sim4\,$arcmin and the background was taken from a region on the same chip. Individual spectra and response files for each detector and data mode were created using the tasks \texttt{xisrmfgen} and \texttt{xissimarfgen}. Finally, the spectra and response files of all data modes for a given detector were summed, weighting by exposure time, to yield one spectrum for each XIS. Light curves were also created from the reprocessed event files using \texttt{xselect}.

The Hard X-ray Detector \citep[HXD;][]{takahashi:07a} data were first reprocessed as recommended by the ABC Guide, and then filtered using \texttt{xselect} with the standard criteria. Only the PIN data is considered in this work as GSO is not suitable for sources with the count rate as low as Fairall~9.   For the PIN, the appropriate ``tuned" background files were downloaded from the High Energy Astrophysics Science Archive Research Center (HEASARC). We then extracted source and background spectra from the ``cleaned" event files using \texttt{hxdpinxbpi}. 

\begin{figure}[t]
\centerline{
\includegraphics[width=0.7\columnwidth]{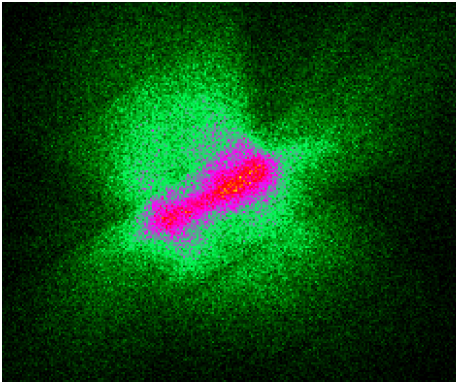}
}
\caption{Image for XIS3 from uncorrected \textit{Suzaku} pointing for Fairall~9.}\label{image}
\end{figure}

In addition to the reduction steps outlined above, the May-2010 Fairall\,9 \textit{Suzaku} data had to be treated with special caution as it is affected by a variety of calibration/pointing problems.    As one can see in Fig.~\ref{image}, the pointing was unstable and the standard attitude solution was inaccurate during this observation, with the satellite wobbling back and forth between two positions separated by 1\,arcmin. Although it is known that \textit{Suzaku} is suffering from an attitude control problem and therefore a mild form of this is already accounted for in the calibration files delivered with the data, it appeared to be too strong to be properly removed with the standard correction (Suzaku memo 2010-05).   The best results are achieved by creating a new, corrected attitude file using the tool \texttt{aeattcor.sl} \footnote{http://space.mit.edu/CXC/software/suzaku/aeatt.html} written by Mike Nowak; a detailed description of this tool can be found in \citet{nowak:11a}.

Once corrected for, the wobble leaves three remaining imprints on the spectra.  Firstly, since the wobble puts some of the source photons into a region of the image plane where the high-energy vignetting function is more poorly known, there are some deviations between the different XIS spectra above 8--9\,keV.  We choose to use XIS data up to 10\,keV but must be cautious about interpreting spectral residuals at the highest XIS energies. Secondly, as is well known, the XIS spectra are contaminated by absorption from a hydrocarbon layer residing on the optical blocking filter.   This contamination is monitored by the {\it Suzaku} team and corrected for in the construction of the effective area file.  However, the contamination has spatial dependence and, for XIS0, it appears that the wobble during this observation has placed substantial numbers of source photons on parts of the image plane in which the contamination is poorly known.  As a result, the XIS0 spectrum deviates from the other XIS spectra below 2\,keV.   This deviation is significant (reaching values as high as 50\,\%) and forces us to ignore XIS0 data below 2.3\,keV.   Lastly, due to the vignetting issue, there is no reason to suspect that the standard XIS/PIN cross normalization factor of 1.18 (nominal HXD aimpoint; Suzaku memo 2008-06) is appropriate for this dataset --- the suppression of XIS flux by the additional vignetting would increase this cross normalization.  We must allow this cross normalization to be a free parameter and, indeed, our best fit models suggest values of 1.3.  

\subsubsection{XMM-Newton}

The two \textit{XMM-Newton} datasets were reduced using the \textit{XMM-Newton} Science Analysis System (SAS) version 10.0.0 and the CCF calibration file XMM-CCF-REL-270.   We only consider EPIC-pn data for the analysis performed in this paper, as MOS does not have enough signal-to-noise in the iron K region to contribute significantly to the constraints.   The EPIC-pn data were reprocessed and screened using the tool \texttt{epchain}. The spectra for the EPIC cameras were then extracted using the tool \texttt{evselect}, selecting the default grade pattern. Source spectra are taken from a circular region and the background spectra from a region on the same chip. The effective area files were generated using the SAS task \texttt{arfgen}, and the redistribution matrices were produced using the task \texttt{rmfgen}.

\section{Results}\label{spectral}

\subsection{Initial Data Exploration and Model Construction}\label{initial}
Unless explicitly stated otherwise, all XIS/EPIC-pn spectra are binned to a signal-to-noise ratio of 10. The considered energy ranges for fitting are 0.7-1.5\,keV and 2.3-10\,keV for \textit{Suzaku}-XIS, with the energies around 2\,keV being excluded because of known calibration issues around the mirror (gold) and detector (silicon) edges. The only exception to this is the XIS0 spectrum for the newest Fairall~9 \textit{Suzaku} pointing; as already mentioned in \S\ref{suz_red} this spectrum suffers from uncorrected contamination at soft energies and, therefore, we ignore the region of this spectrum below 2.3\,keV.  EPIC-pn spectra are used in the 0.5-10\,keV band. For all \textit{Suzaku} pointings, PIN spectra were binned to S/N of 5 and are used in the 16-35\,keV band.   As our spectral models include a convolution with a relativistic transfer function, which requires an evaluation of the underlying model outside of the energy range covered by the data, we extended all energy grids to energies far beyond the upper energy limit given by the highest data bin considered in fitting.   The observations considered in this work span a duration of 10\,years (Table~\ref{tab1}), and hence give an idea of the long-term variability of the object. None of these observations show significant short-term (intra-observation) variability. Therefore, our spectral analysis considers only pointing averaged spectra.

\begin{figure}[t]
\includegraphics[width=\columnwidth]{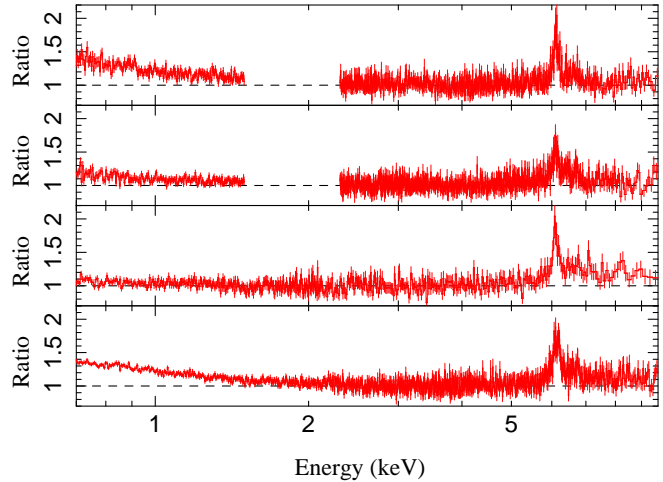}
\caption{Ratio to a simple power law for Fairall~9 for Suzaku A [Top Panel], Suzaku B [Upper Middle Panel], XMM A [Lower Middle Panel], XMM B [Bottom Panel].}\label{f9_data}
\end{figure}

\begin{figure}[t]
\includegraphics[width=\columnwidth]{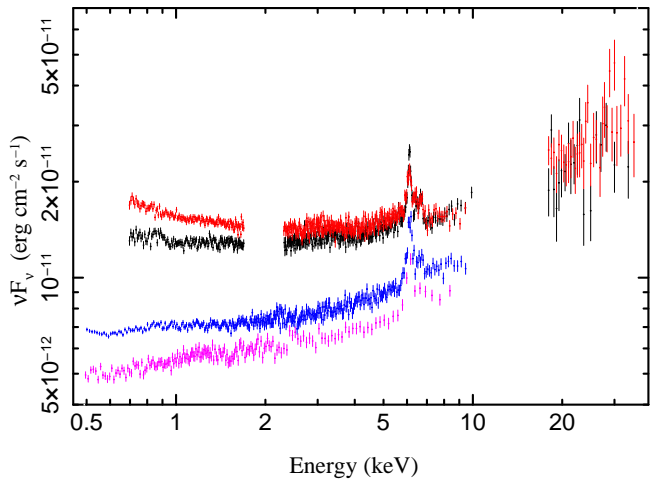}
\caption{Unfolded spectra (Suzaku A [black], Suzaku B [red], XMM A [purple], XMM B [blue]) of Fairall~9 showing the spectral variability of the source. The spectra were rebinned for plotting.}\label{un}
\end{figure}

As an initial exploration of these data, we fit a simple absorbed power-law to the 2.3--4.5\,keV part of the spectrum; Figure~\ref{f9_data} shows the ratio of the data in the 0.5--10\,keV band to this simple model for all four datasets.  All datasets show a prominent narrow, fluorescent iron K$\alpha$ line (6.4\,keV rest frame) as well as the blend of the K$\beta$/Fe XXVI lines (7.07\,keV/6.96\,keV rest frame).    The spectrum shows clear concavity, revealing the presence of a soft excess and, possibly, a Compton reflection.  The two narrow fluorescent iron lines together with the reflection hump are clear evidence for cold reflection in the source, either from the torus or from the outer parts of the accretion disk. None of the spectra show any signs of warm absorption, as already discussed in \S\ref{intro}. Despite these similarities, the spectral shape varies significantly throughout the years, with the \textit{XMM-Newton} pointings (2000, 2009) catching the source in a lower flux state. This variability is even more apparent when plotting the unfolded\footnote{The unfolding is performed by assuming a diagonal response matrix not making any assumptions about the description of the spectra.} spectra (Fig.~\ref{un}).  As the signal-to-noise in the short (2000) \textit{XMM-Newton} pointing is significantly lower than the other datasets we do not consider this pointing for the remaining part of this work.   

Building upon previous works \citep{schmoll:09a,emmanoulopoulos:11a}, we construct a multi-component spectral model to describe our spectra.     The underlying primary continuum is described by a powerlaw with photon index $\Gamma$.   We use the model \texttt{pexmon} \citep{nandra:07a} to describe the cold reflection of this powerlaw from distant material.   In particular, \texttt{pexmon} self consistently models the strength of the narrow Fe~K$\alpha$, Fe~K$\beta$, and Ni~K$\alpha$ lines, the Compton shoulder of the Fe K$\alpha$ line, and the Compton reflection continuum.  The strength of the distant reflection is characterized by the usual reflection fraction $R$ normalized such that $R=1$ corresponds to a reflector that subtends half of the sky as seen from the X-ray source. As some parameters for \texttt{pexmon} cannot be determined from our fits we fix them to certain values; we fix the inclination to 60\,degrees (most probable value for an isotropic distribution) and abundances to solar values.   The high-energy cutoff of the continuum powerlaw is fixed to 300\,keV.

To describe the ionized reflection associated with the inner accretion disk, we used a modified version of \texttt{reflionx} developed by \citet{ross:05a}. Starting from the publicly available version, we redefined the normalization parameter (${\rm norm}\rightarrow {\rm norm}/\xi$) such as to statistically decouple the true flux normalization from the ionization parameter $\xi$.   This definition results in a more rapid and robust convergence of the spectral fit.  For the ionized reflection model, the iron abundance and ionization parameter are allowed to vary freely. The photon index of the irradiating continuum, however, is tied to the index for the continuum underlying the cold reflection. The ionized reflection component is then relativistically blurred using the model \texttt{relconv} \citep{dauser:10a}; this naturally gives rise to a relativistic iron line, blurred Compton reflection hump and a soft excess. The radial dependence of the emissivity of the reflection component is assumed to have a broken power-law form, breaking from $r^{-q_1}$ to $r^{-q_2}$ at a radius $R_{\rm break}$.    The inner edge of the X-ray reflection regime is taken to be at the ISCO \citep{reynolds:08a}, and the outer edge was fixed to 400\,$R_\text{g}$.  Provided that $q_2>2$, the relativistic blurring kernel is only weakly dependent upon this outer radius.   The accretion disk inclination $i$, and the black hole spin $a$ were left as free parameters.   In addition, photoelectric Galactic absorption is modeled with \texttt{TBnew}\footnote{http://pulsar.sternwarte.uni-erlangen.de/wilms/research/tbabs/} a newer version of \texttt{TBabs} \citep{wilms:00a} with cross sections set to \texttt{vern} and abundances set to \texttt{wilm}. 

\begin{figure}[h]
\centerline{
\begin{minipage}{0.48\textwidth}
\includegraphics[width=\textwidth]{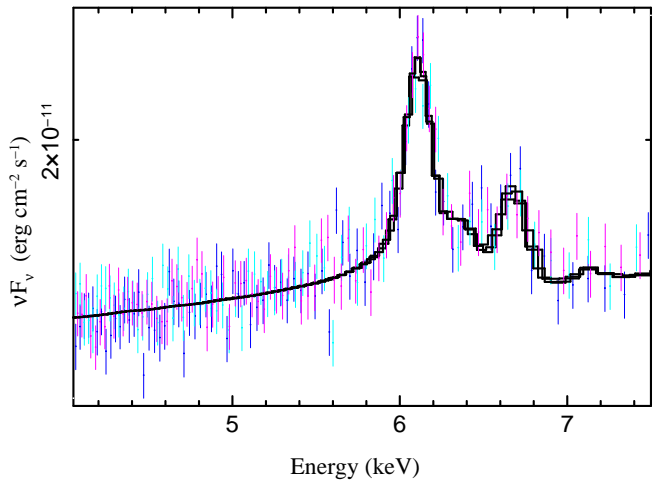}
\caption{Unfolded spectra (rebinned for plotting), assuming a diagonal response, and model [black] of the iron line region for observation Suzaku-A fitted with the model including photoionized emission (see \S\ref{initial}). The plot shows the energies in the observed frame.}\label{f9_line}
\end{minipage}
}
\end{figure}

A careful examination of the residuals shown in Fig.~\ref{f9_data}, especially for observation Suzaku-B, reveals a line like feature between cold-FeK$\alpha$ and cold-FeK$\beta$ lines.   While some or all of this emission may be associated with the blue peak of the broad iron line, another possibility is that we are seeing FeK$\alpha$ line emission from Fe\,XXV (which produces a complex at 6.7\,keV).  It therefore seems possible that the spectral model discussed above is missing an emission line system producing Fe\,XXV and possibly Fe\,XXVI line emission.  Hence we include an additional photoionized emission component (Fig.~\ref{f9_line}); physically, this may be an accretion disk wind (warm absorber) which is out of our line of sight. The photoemission is parametrized by ionization parameter and norm. It is modeled using a table model calculated using \texttt{xstar2xspec} from the \texttt{XSTAR} model \texttt{photemis}\footnote{http://heasarc.nasa.gov/docs/software/xstar/xstar.html}, assuming irradiation by a power law $L_\varepsilon \propto \varepsilon^\alpha$ with $\alpha$=-1 and a number density of $10^{10}\,\text{cm}^{-3}$.   At the ionization stages relevant for this work ($\log\xi>3.6$) there is no dependence of the spectral shape on column density for the model, its value was therefore kept fixed at $10^{22}$\,cm$^{-3}$.   We note that the inclusion of this emission line component this does not preclude the possibility that the residuals at 6.7\,keV are from the blue peak of the iron line, but it does allow the model to explore any degeneracy created by the superposition of the broad iron line and the narrow ionized line emission.  

When applying this spectral model to the data, a cross calibration constant was introduced for \textit{Suzaku} between XIS and PIN (and in case of the 2010 \textit{Suzaku} pointing, also between the XISs). The constant was fixed to the values given in the \textit{Suzaku} ABC Guide, 1.16 for XIS aimpoint and 1.18 for HXD aimpoint. An exception to this is observation Suzaku-B; as mentioned above, due to the pointing issues, there is no expectation that the fiducial XIS/PIN cross normalization should be appropriate and hence it was left as a free parameter. 

We begin with this model as our base model and, in \S\ref{fit}, we fit this model to each dataset individually.  After noting problematic issues associated with these fits, we perform simultaneous fits to data from different observations of Fairall~9, under the assumption that the most fundamental parameters of the system (spin, accretion disk inclination and iron abundance) do not change --- this so-called multi-epoch fitting is presented in \S\ref{multi} and, as we shall see, yields insights into the spectral properties of this AGN.   

Throughout this work, the spectral analysis were performed with the Interactive
Spectral Interpretation System\footnote{http://space.mit.edu/cxc/isis} \citep[ISIS Version 1.6.0-7;][]{houck:00a} and the newest \textsl{XSPEC} 12.0 models are used \citep{arnaud:96a}. All uncertainties are quoted at the 90\,\% confidence level for one interesting parameter ($\Delta\chi^2=2.7$).

\subsection{Independent fits to individual datasets}\label{fit}

\begin{figure}[t]
\centerline{
\begin{minipage}{0.48\textwidth}
\includegraphics[width=\textwidth]{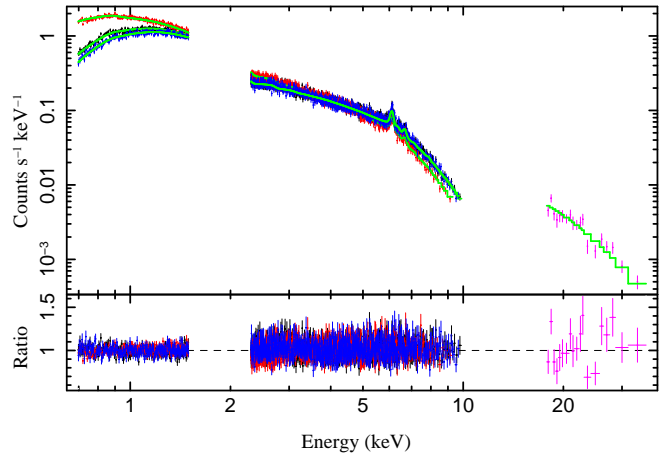}
\caption{Spectra, model [green] and residuals for pointing Suzaku-A (XIS0 [black], XIS1 [red], XIS3 [blue], PIN [purple]) for the fit with the base model.}\label{f9_cbase}
\vspace*{2\baselineskip}
\end{minipage}
}
\end{figure}

We begin by conducting independent fits of our base spectral model to the three remaining observations under consideration, XMM-Newton-B, Suzaku~A and Suzaku~B.    Our base model provides a good description of data well for three datasets, with reduced-$\chi^2$ values between 1.00--1.04 (Table~\ref{table_base}). A sample spectrum with residuals is shown in Fig.~\ref{f9_cbase} illustrating the good quality of the fit.

We argue below that these fits are incomplete and hence that the inferred parameters are thus not necessarily correct.  However, it is instructive to compare our results from these fits (Table~\ref{table_base}) with previously published works.   For Suzaku~A, our derived parameters are similar to those found by \citet{schmoll:09a} with the exception of black hole spin, disk inclination, and emissivity index.   \citet{schmoll:09a} apply the constraint that the inner emissivity index $q_1$ should not exceed 5; if we impose the same constraint, we recover a very similar intermediate spin and inclination constraint.  However, our base model allows the emissivity index to be much steeper (formally pegging at $q_1=10$) and we find,as a result, that the inferred black hole spin tends to high values.      For Suzaku~B, our fits are in agreement with \citet{patrick:11a} who finds that spin is unconstrained by this dataset.   For XMM-Newton~B, we find a spin parameter and inclination that is strongly discrepant with those derived in \citet{emmanoulopoulos:11a}.  We attribute this difference to the inclusion of a photoemission component, a newer absorption model (\texttt{TBnew} vs. \texttt{wabs}) and a different model construction.    
 
However, just looking at the results of our own uniform analysis, we find important and informative inconsistencies. Comparing the fit parameters derived from the different observations we find that the black hole spin, the disk inclination, and the iron abundance appear to have significantly different values between the three datasets (Table~\ref{table_base}). Naturally we would expect these physical quantities to be constant on the time spanned by these observations.

As we show below, a multi-epoch analysis reveals the need for a new continuum component to describe the soft excess.   Foreshadowing that discussion, we note that adding such a soft excess component to the model for each individual dataset results in unconstrained spin parameters and emissivity indices, i.e., the individual datasets do not possess the S/N to determine spin and inclination in the more complex models; there is a significant trade off between the parameters of the soft component and the spin and inclination for each individual data set. This also implies that that the extreme values of spin and $q_1$ found in a fit of the base model to Suzaku-A may arise due to a model which is (possibly falsely) attempting to fit a very smooth soft excess with an ionized reflection spectrum, necessitating extreme broadening.    

\subsection{Multi-epoch analysis}\label{multi}

The inconsistencies resulting from independently fitting our base model to the individual pointings leads us to the use of multi-epoch fitting.    In multi-epoch fitting, we assume that the source spectrum is always composed of the same principal physical components and that, when physical considerations demand, parameters are forced to have the same value for all epochs.   In our case, we demand that the black hole spin parameter, the inclination of the disk, the iron abundance, and the strength of the photoionized emission component (believed to originate from an extended AGN wind) must have common values across the fits to all of the datasets.   In all cases, these parameters are believed to remain fixed for any given AGN over human timescales.   Aside from these additional constraints on the parameters, the model set-up is the same as for the fitting of the individual pointings.   We allow the slope/normalization of the primary powerlaw, as well as the normalization, emissivity profile and ionization state of the disk reflection to float freely between datasets --- these parameters depend upon the structure/geometry of the disk corona that can plausibly change on timescales that are (much) shorter than the inter-pointing spacing.    Fitting in this manner we find that the broken power law emissivity profile is ill-constrained and a simple power law emissivity profile is sufficient to describe the data very well.  Furthermore, to reduce the number of spectral bins in each dataset, we co-added the front-illuminated XIS data (XIS0 and XIS3); this eases the significant computational expenses of multi-epoch fitting.   

Statistically, fitting this model to the full multi-epoch dataset produces a good fit, with reduced $\chi^2$ of 1.06. The best fit parameters and errors are listed in Table~\ref{mtable_base}.   The best fitting spin is a little higher ($a=0.71^{+0.08}_{-0.09}$) and the inclination a little lower ($i=37^{+4}_{-2}$\,deg) than found in the previous study of \cite{schmoll:09a} but, given the error bars, there is no strong discrepancy.   The ionization parameter either takes rather low values (XMM, Suzaku A) or very high values (Suzaku B).   This is directly tied to the fact that, apart from the iron line itself, there are no other strong features in the soft X-ray band that the reflection models are locking onto in the spectrum.   One aspect of this fit that gives us pause is the high PIN/XIS cross normalization component, $C_{\rm XISf-PIN}\approx 1.74$.   While we have already acknowledged that the pointing errors would render invalid the standard value of $C_{\rm XISf-PIN}=1.18$, this measured value would imply a $\sim 30$\% suppression of the net XIS count rate below the fiducial value, a significantly larger depression than is reasonably expected from the pointing problems.  A more likely possibility is that the spectrum has curvature in the concave sense so that the true PIN-band flux is higher than predicted by our base spectral model.   

Partly motivated by these considerations, but also guided by the known phenomenology of AGN, we explore the hypothesis that we are missing another component that can contribute to the soft excess (above and beyond any soft excess emission from the blurred uniformly ionized reflection, that models the broad iron line).  A common phenomenological model for the soft excess is a simple blackbody component.  Adding a blackbody component to the spectral model does not yield improved fits --- none of the datasets show the strong curvature representative of the Wien tail of a blackbody.   However, significant improvements in the goodness of fit are achieved using a soft excess model consisting of a ``warm" thermal Comptonization component.   In detail, we add a thermal Comptonization component described by the {\tt comptt} model \citep{titarchuk:94a} with a seed photon temperature of 40\,eV (representative of the expected thermal emission from the AGN disk), an optical depth $\tau$ and an electron temperature $kT$.   

When applied to the multi-epoch data, the inclusion of this soft excess component leads to an improvement in the goodness of fit by $\Delta \chi^2=198$ for nine additional parameters.   The best fit parameters and corresponding errors can be found in Table~\ref{mtable_soft}.  As compared with the base model above, the best-fit spin has slightly decreased to $a=0.52^{+0.19}_{-0.15}$ and the best-fit inclination has increased to $i=48^{+6}_{-2}$\,deg.   While the change in spin between the two models is within the error bars, the inclination change is significant and brings the results  into line with the conclusions of \citet{schmoll:09a}.  We find that the ionization state of the inner accretion disk is significantly higher at lower flux states.   The biggest surprise is that, once we include the explicit soft excess component, the iron abundance of the accretion disk is pushed towards the maximum value tabulated in the model, $Z=10Z_\odot$.     The formal 90\% lower limit on iron abundance is $Z>8.2Z_\odot$. From a modeling perspective, such a high iron abundance means simply that the iron line is strong in comparison to the Compton hump. 


However, another possibility is that the reflection from the inner accretion disk is characterized by multiple ionization components \citep{nardini:11a} and a highly blurred, high-ionization reflection component can account for the soft excess in a similar fashion than the additional Comptonization component.  To explore this model, we refit the multi-epoch data with two disk reflection components, each of which has its own ionization parameter and emissivity index. The only other difference to the multi-epoch fit above is that the iron abundance is assumed to be the same for all reflection components included in the model.  Such a model leads to a fit solution of almost comparable quality than that employing the Comptonization-based soft excess component (Table~\ref{mtable_double}).  In order to describe the smooth soft excess, one of the disk reflection components adopts a high ionization state ($\xi_a\sim 1000$) and a very steep emissivity profile ($q_a>8$) and the black hole spin parameter becomes large ($a>0.96$). At the same time, much less broadening/blurring is needed in order to describe the broad iron line, and so the other disk reflection component adopts a low ionization ($\xi_b\sim 1-10$) and a shallow emissivity profile ($q\lesssim 2.2$). This solution avoids the extreme iron abundance, with an inferred iron abundance of 0.68--0.85\,$Z_\odot$.

\section{Summary and discussion}\label{dis}

The X-ray spectrum of Fairall~9 is one of the cleanest known in terms of having no discernible intrinsic absorption.   Fairall~9 has also been the subject of multiple pointings by {\it XMM-Newton} and {\it Suzaku}.    This makes it an interesting target for a detailed study of the relativistic reflection features including an assessment of the robustness of the derived parameters with respect to assumptions and choices made in the spectral modeling. 

\begin{figure*}[t]
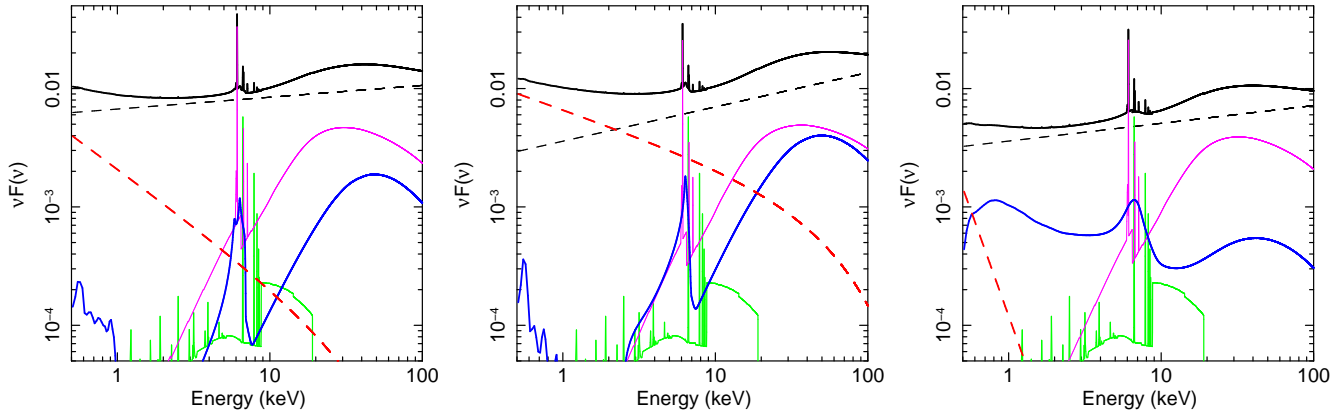

\includegraphics[width=0.3\textwidth,angle=270]{fig6a.eps}
\includegraphics[width=0.3\textwidth,angle=270]{fig6b.eps}
\includegraphics[width=0.3\textwidth,angle=270]{fig6c.eps}
\caption{Best fitting multi-epoch model (including the additional soft continuum source) for the Suzaku~A (left), Suzaku~B (middle) and XMM-Newton~B (right) observations.    Shown here is the total model (solid-thick black line), the primary power-law continuum (dashed black line), warm Comptonization component (dashed red line), the relativistically blurred ionized disk reflection (thick solid blue line), the distant reflector (thin solid magenta line), and the photoionized emission component (green line).}\label{f9_multispec}
\end{figure*}

A simple application of our base spectral model to the individual {\it XMM-Newton} and {\it Suzaku} pointings produces unphysical results, with inconsistent values for black hole spin, disk inclination and iron abundance being derived for each pointing.    We identify several factors that are at play.   We clearly see (and hence include in our base model) high ionization, narrow iron emission lines, presumably from optically-thin photoionized plasma in the circumnuclear environment.  When viewed with CCD-resolution ($E/\Delta E\sim 50-100$), these narrow lines can blend with the blue wing of the broad iron line from the accretion disk, inducing strong degeneracies between the strength of the photoionized emission component and the accretion disk parameters.  This degeneracy is particularly strong for objects with intermediate inclination (35--60\,deg) since, in these cases, the blue wing/edge of the iron line overlays the FeXXV K$\alpha$ line complex.

To cut through this degeneracy, we make the assumption that the photoionized emission component is constant over the three years that span observations {\it Suzaku}-A, {\it XMM-Newton-B}, and {\it Suzaku}-B.   We then employ a multi-epoch analysis, fitting our spectral model to these three datasets simultaneously --- the black hole spin, accretion disk inclination, iron abundance, and photoionized emission component are assumed constant across all three pointings, but all other spectral parameters and normalizations are allowed to float freely between pointings.  Since there is, indeed, significant flux variability between the three pointings, this approach is remarkably successful at breaking the degeneracy with the photoionized emission component and allowing us to constrain model values for the black hole spin and disk inclination.   Of course, it must be highlighted that the constancy of the photoionized emission component is an {\it assumption} which could be violated if the emitting wind is compact (sub-parsec).   This assumption must be tested by future high-resolution observations using the {\it Chandra} High Energy Transmission Gratings (HETG) or the {\it Astro-H} Soft X-ray Spectrometer (SXS).

In addition, we find that the multi-epoch spectral fit is significantly improved by the addition of another spectral component. Our analysis therefore highlights the practical importance of modeling the soft excess.   The nature of the soft excess is still a matter of debate, and the soft Comptonization component that we attempt first is just one of the possible models discussed in the context of AGN spectra.   In the case of Fairall~9, we have already excluded a soft excess that arises solely from inner disk reflection with a single ionization parameter, and have shown that an additional blackbody component is also rejected.  

We first consider the scenario in which the soft excess is identified with an additional Comptonization component. As explicitly shown in Fig.~\ref{f9_multispec}, this additional Comptonization component is the main contributor to the soft excess during our two Suzaku observations while ionized disk reflection dominates during the XMM-Newton~B pointing.    The inclusion of this soft continuum has a small effect on the best fitting black hole spin and disk inclination (with our final values being $a=0.52^{+0.19}_{-0.15}$ and $i=48^{+6}_{-2}$\,deg), but has a dramatic effect on the inferred iron abundance of the accretion disk ($Z>8.2Z_\odot$). With an expected abundance value of only 2-3$Z_\odot$ \citep{groves:06a}, such a high iron abundance implies an unusual star formation history in the galactic core, or some process that preferentially differentiates iron into the photosphere of the inner disk (e.g., see discussion of Reynolds et al. 2012). 

It is interesting that, in our fiducial multi-epoch fit (Table~\ref{mtable_soft}), the {\it Suzaku}-B dataset requires a high emissivity index, $q>6$.   Such high emissivity indices are normally attributed to the action of extreme light bending focusing X-rays onto the innermost regions of the disk or, alternatively, the dissipation of work done by strong torques at the ISCO.  Both of these explanations would require a black hole that is spinning rapidly ($a>0.9$), as opposed to the intermediate spin ($a\approx 0.5$) that we infer \citep{fabian:12a}.  A possible resolution of this contradiction comes from the vertical geometry of the accretion disk.  Fairall~9 is a luminous source accreting with an Eddington ratio of ${\cal L}\approx 0.15$ \citep{lohfink:12a}.   Standard disk theory \cite{shakura:73a} tells us that the inner regions of the accretion disk will be very radiation pressure dominated and, away from the inner boundary, will have a disk thickness $h=(3{\cal L}/2\eta)r_g\approx 2.3r_g$ where we have taken the radiative efficiency of the disk to be $\eta=0.1$.   As the disk approaches the ISCO (at $r_{\rm isco}\approx 4.5r_g$ for $a=0.5$) the thickness diminishes and, within the ISCO, the accretion flow forms a thin sheet that spirals into the black hole.   This gives the photosphere of the inner disk the geometry of shallow bowl.  If the X-ray source is located very close to the black hole (e.g. in a spin-powered magnetosphere) or on the inward facing surface of this bowl, the region close to the ISCO can be strongly irradiated while the disk surface at large distance may receive very little irradiation --- this would manifest itself in our spectral analysis as a steep emissivity profile. Alternatively, the high emissivity index may be a mirage induced by a steep iron abundance gradient. Specifically, if the high iron abundance is due to radiative-levitation in the disk photosphere \citep{reynolds:12a}, it will have a strong radial gradient which would appear in our fits as a steep emissivity index.  

\begin{figure*}[t]
\includegraphics[width=0.45\textwidth]{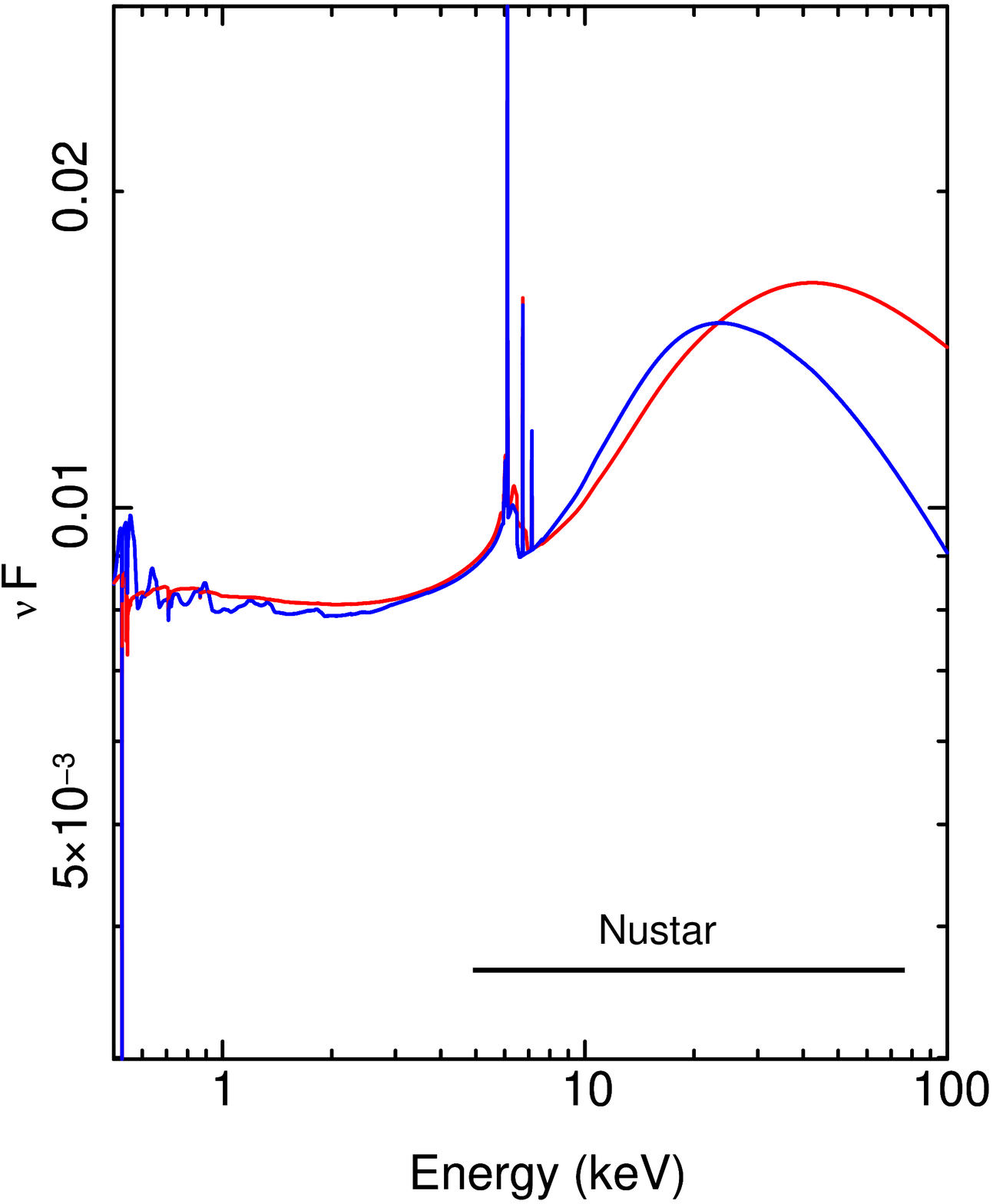}
\includegraphics[width=0.45\textwidth]{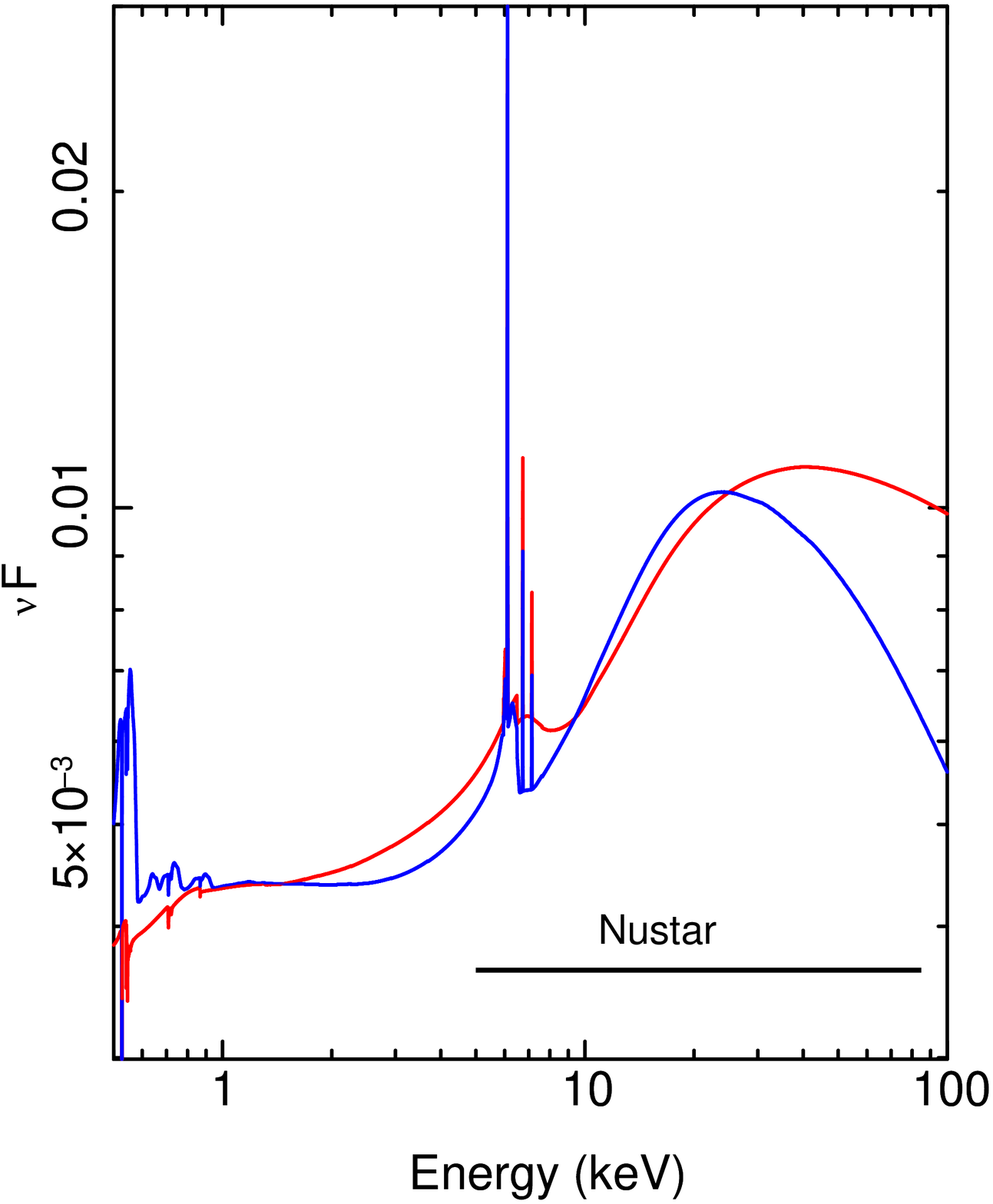}
\caption{A comparison of the two best fitting spectral models, the double reflection model (blue) and the one with additional Comptonization soft excess (red), for the first \textit{Suzaku} pointing (left) and the \textit{XMM} pointing (right). The models differ significantly in the \textit{Nustar} energy range marked with the black bar.}\label{f9_highspec}
\end{figure*}

Alternatively, this steep emissivity profile could be the artifact of incorrect modeling of contamination in the spectrum. As outlined in \S\,\ref{suz_red}, this \textit{Suzaku} pointing is certainly affected by strong contamination in XIS0 below 1\,keV. We assumed for this analysis that the other two detectors were correct below 1\,keV, but the steep profile could be an indication that this assumption is not justified.

The alternative scenario that we consider for the soft excess in Fairall~9 is a model consistent of two ionized reflection components with different ionization states. This double-ionized reflection solution does present difficulties, however, when we attempt to interpret it with physical models.   There are two classes of model in which multiple ionization states may be present in the disk reflection spectrum.   Firstly, at any given radius, the density of the disk photosphere may be patchy leading to mixed regions of high and low ionization.   In such a scenario, we would expect the two ionization components to share approximately the same emissivity index since they respond to the same irradiation profile.   This disagrees with the fact that we see very different emissivity indices between these two components and allows us to reject this patchy disk model.   Secondly, we may expect the ionization parameter of the disk surface to have a strong radial gradient due to the radial dependence of both the irradiating flux and the disk density.   The fact that we see a high-ionization component which is highly blurred (coming from the smallest radii) and a low-ionization component which is more weakly blurred (coming from substantially larger radii) fits in with this picture.  However, preliminary theoretical calculations of the reflection from a disk with a radial ionization gradient reveals that the intermediate ionization states should also imprint strong soft X-ray features which we do not see \citep{garcia:12a}.

Fundamentally, however, it is difficult to distinguish these two scenarios (new Comptonization continuum or second disk reflection component) with existing data. Although the models do not differ significantly in quality of their description of the data, the physical interpretations are quite distinct. Plotting the models to energies well above the PIN energy range  (Fig.~\ref{f9_highspec}) points out how future hard X-ray studies with \textit{Nustar} or \textit{Astro-H} could break this degeneracy. The hard X-ray flux predicted by the double reflection scenario is substantially smaller than the one predicted by the model including the additional Comptonization component. In addition, the high spectral resolution will help in the soft X-ray part putting firm constraints on the smoothness of the X-ray excess.

Fairall~9, with its absence of warm absorption, is an interesting target to study the properties of the soft excess. This enables us to test possible scenarios explaining the soft excess in the spectrum, as well as the time variability of this component. We note that the shape of the additional soft excess seems to change as the source becomes dimmer (Fig~\ref{f9_multispec}). The dimmer the source the steeper the additional component. At the time the contribution of the additional component decreases and the lower ionized reflection dominates. A more detailed analysis of this phenomenology could lead to more insight into the nature of this additional component but requires more high quality pointings and is therefore subject to future studies. 

As a final comment, we highlight that the {\it Suzaku}-B pointing needs to treated carefully and the cross normalization between XIS and HXD-PIN cannot be expected to be at normal value but instead is found to be higher (1.3). We note that the reported discovery by \cite{tatum:12a} of a new spectral component that only reveals itself in the PIN-band of this particular dataset (and interpreted as a transitory marginally Compton-thick absorber) may be affected by this incorrect cross-normalization.



\acknowledgments
\section*{Acknowledgments}

We thank the \textit{Suzaku} Guest Observer Facility members for their help with the newest \textit{Suzaku} pointing. The authors thank the anonymous referee for comments that improved this manuscript. This research has made use of: data obtained from the High Energy Astrophysics Science Archive Research Center (HEASARC) provided by NASA's Goddard Space Flight Center; NASA's Astrophysics Data System Bibliographic Services, observations obtained with \textit{XMM-Newton}, an ESA science mission with instruments and contributions directly funded by ESA Member States and the USA (NASA).     AML and CSR acknowledges support from NASA through the Suzaku Cycle-5 Guest Observer Program (grant NNX10AR31G) and the Astrophysical Data Analysis Program (grant NNX12AE13G).

Facilities: \facility{\textsl{Suzaku}}, \facility{\textsl{XMM-Newton}}


\clearpage
\newpage

\begin{table}[ht]
\begin{center}
\caption{Spectral Parameters for Fairall~9 base model fits with additional photoionized emission; See the text for a description of the model. \textit{Suzaku} Spectra are normalized to XIS0 data. The power law normalization is $\text{photons}\,\text{keV}^{-1}\,\text{cm}^{-2}\,\text{s}^{-1}$ at 1\,keV.\label{table_base}}

\begin{tabular}{cccccc}
\hline 
& & Suzaku A & Suzaku B & XMM B \\
 \hline \hline
contiuum \& & $A_\text{pex}$ [$10^{-3}$] & 8.60$_{-0.08}^{+0.08}$ & 10.49$_{-0.08}^{+0.01}$ & 4.80$_{-0.02}^{+0.02}$ \\
cold reflection & $\Gamma$ & 2.11$_{-0.02}^{+0.01}$ & 2.11$_{-0.01}^{+0.01}$ & 2.11$_{-0.02}^{+0.03}$ \\
 & $R$ & 1.08$_{-0.11}^{+0.11}$ & 0.76$_{-0.09}^{+0.09}$ & 1.53$_{-0.13}^{+0.14}$ \\
 \hline
ionized reflection &$A_\text{reflionx}$ [$10^{-4}$] & 2.59$_{-0.20}^{+0.39}$ & 1.74$_{-0.22}^{+0.22}$ & 2.00$_{-0.31}^{+0.29}$ \\
 & $Fe/Solar$ & 1.30$_{-0.44}^{+0.16}$ & 1.35$_{-0.49}^{+0.61}$ & 0.50$_{-0.17}^{+0.08}$\\
& $\xi$ [erg\,cm\,s$^{-1}$] & 12.1$_{-1.2}^{+7.2}$ & 1$_{-0}^{+0.01}$ & 1$_{-0}^{+1.34}$\\
\hline
relativistic blurring & $q_1$ & 10$_{-0.6}^{+0}$ & 10$_{-8.1}^{+0}$ & 9.1$_{-1.9}^{+0.9}$ \\
 &$q_2$ & 1.7$_{-0.3}^{+0.5}$ & 1.4$_{-0.9}^{+0.4}$ & 2.4$_{-0.8}^{+0.6}$\\
& $R_\text{break}$ [$R_\text{g}$] & 4.6$_{-0.5}^{+2.1}$ & 8.3$_{-4.5}^{+83.5}$ & 6.5$_{-1.4}^{+0.2}$ \\
 &$a$ & 0.96$_{-0.02}^{+0.01}$ & 0.28$_{-1.28}^{+0.72}$ & 0.93$_{-0.01}^{+0.02}$\\
 & $i$ [deg] & 35.0$_{-5.6}^{+3.4}$ & 62.5$_{-2.4}^{+5.5}$ & 5$_{-0}^{+9.9}$\\
 \hline
plasma & $A_\text{phot}$ [10$^{-3}$] & 0.98$_{-0.73}^{+16.1}$ & 11.47$_{-4.04}^{+52.04}$ & 65.2$_{-15.2}^{+30990.48}$ \\
 &$\log\xi$ & 4.3$_{-0.4}^{+0.5}$ & 3.7$_{-0.0}^{+0.6}$& 4.4$_{-0.1}^{+0.2}$ \\ 
 \hline
cross calibration & $c_\text{XIS0-1}$ & 1.00 & 0.97$_{-0.01}^{+0.01}$ & \nodata \\
 &$c_\text{XIS0-3}$ & 1.00 & 0.92$_{-0.01}^{+0.01}$ & \nodata \\
& $c_\text{XIS0-PIN}$ & 1.16 & 1.43$_{-0.07}^{+0.07}$ & \nodata \\
\hline
 & $\chi^2$/dof & 2593.1/[2500-13] & 2538.1/[2559-16] & 1129.8/[1110-13]\\
 & p-value & 0.07 & 0.52 & 0.25 \\
 \hline
 & $\chi^2_\text{red}$ & 1.04 &1.00 & 1.02\\
 \hline\hline
\end{tabular}
\end{center}
\end{table}

\begin{table}[!ht]
\begin{center}
\caption{Spectral Parameters for Fairall~9 multi-epoch base model fit with additional photoionized emission; See the text for a description of the model. \textit{Suzaku} Spectra are normalized to XIS0 data. The power law normalization is $\text{photons}\,\text{keV}^{-1}\,\text{cm}^{-2}\,\text{s}^{-1}$ at 1\,keV.\label{mtable_base}}

\begin{tabular}{ccccc}
\hline 
& & Suzaku A & Suzaku B & XMM B \\
 \hline \hline
contiuum \& & $A_\text{pex}$ [$10^{-3}$] & 8.66$_{-0.07}^{+0.13}$ & 7.52$_{-1.10}^{+0.57}$ & 4.02$_{-0.07}^{+0.07}$ \\
cold reflection & $\Gamma$ & 2.07$_{-0.01}^{+0.02}$ & 2.05$_{-0.01}^{+0.01}$ & 1.93$_{-0.01}^{+0.01}$ \\
 & $R$ & 0.96$_{-0.08}^{+0.08}$ & 0.95$_{-0.09}^{+0.14}$ & 1.47$_{-0.12}^{+0.12}$ \\
 \hline
ionized reflection &$A_\text{reflionx}$ [$10^{-4}$] & 1.51$_{-0.15}^{+0.28}$ & 0.53$_{-0.09}^{+0.09}$ & 0.34$_{-0.03}^{+0.03}$\\
 & $Fe/Solar$ & \multicolumn{3}{c}{\textbf{0.67}$_{-0.08}^{+0.08}$} \\
& $\xi$ [erg\,cm\,s$^{-1}$] & 6.1$_{-3.8}^{+3.2}$ & 1739.2$_{-509.2}^{+1142.6}$ & 500.0$_{-107.9}^{+18.5}$ \\
\hline
relativistic blurring & $q$ & 2$_{-0}^{+0.23}$ & 2.57$_{-0.40}^{+0.47}$ & 8.61$_{-1.88}^{+1.39}$ \\
  &$a$ & \multicolumn{3}{c}{\textbf{0.71}$_{-0.09}^{+0.08}$}\\
 & $i$ [deg] & \multicolumn{3}{c}{\textbf{37}$_{-2}^{+4}$}\\
 \hline
plasma & $A_\text{phot}$ & \multicolumn{3}{c}{\textbf{24.85}$_{-24.77}^{+46.51}$}
\\
 &$\log\xi$ & \multicolumn{3}{c}{\textbf{6.7}$_{-2.4}^{+0.3}$} \\ 
 \hline
cross calibration & $c_\text{XISf}$ & 1.00 & 1.00 & \nodata \\
 &$c_\text{XIS1}$ & 1.00 & 0.99$_{-0.01}^{+0.01}$ & \nodata \\
& $c_\text{XISf-PIN}$ & 1.16 & 1.74$_{-0.08}^{+0.08}$ &\nodata \\
\hline
individual & $\chi^2$/dof & 2228.29/[2106-11] & 2071.272/[2060-13] & 1447.278/[1100-11] \\
contributions & p-value & 0.02 & 0.35 & 0.00 \\\hline \hline
 & $\chi^2$/dof & \multicolumn{3}{c}{5544.218/[5276-25]} \\
  & p-value & \multicolumn{3}{c}{0.002} \\
 & $\chi^2_\text{red}$ & \multicolumn{3}{c}{1.06} \\
 \hline\hline
\end{tabular}
\end{center}
\end{table}

\begin{table}[!ht]
\begin{center}
\caption{Spectral Parameters for Fairall~9 multi-epoch fit with additional photoionized emission and soft excess component; See the text for a description of the model. \textit{Suzaku} Spectra are normalized to XIS0 data. The power law normalization is $\text{photons}\,\text{keV}^{-1}\,\text{cm}^{-2}\,\text{s}^{-1}$ at 1\,keV.\label{mtable_soft}}

\begin{tabular}{ccccc}
\hline 
& & Suzaku A & Suzaku B & XMM B \\
 \hline \hline
contiuum \& & $A_\text{pex}$ [$10^{-3}$] & 6.71$_{-0.78}^{+0.53}$ & 3.59$_{-1.62}^{+1.55}$ & 3.60$_{-0.18}^{+0.17}$ \\
cold reflection & $\Gamma$ & 1.90$_{-0.05}^{+0.04}$ & 1.71$_{-0.09}^{+0.15}$ & 1.85$_{-0.02}^{+0.02}$ \\
 & $R$ & 0.84$_{-0.07}^{+0.07}$ & 0.79$_{-0.17}^{+0.13}$ & 1.08$_{-0.11}^{+0.11}$ \\
 \hline
ionized reflection &$A_\text{reflionx}$ [$10^{-4}$] & 0.57$_{-0.12}^{+0.12}$ & 0.99$_{-0.18}^{+0.34}$ & 0.13$_{-0.02}^{+0.02}$\\
 & $Fe/Solar$ & \multicolumn{3}{c}{\textbf{10}$_{-1.77}^{+0}$} \\
& $\xi$ [erg\,cm\,s$^{-1}$] & 20.9$_{-8.8}^{+14.8}$ & 10.1$_{-9.1}^{+60.4}$ & 3513.5$_{-393.7}^{+381.0}$ \\
\hline
relativistic blurring & $q$ & 2$_{-0}^{+0.23}$ & 9.5$_{-3.4}^{+0.5}$ & 2$_{-0}^{+0.33}$ \\
  &$a$ & \multicolumn{3}{c}{\textbf{0.52}$_{-0.15}^{+0.19}$}\\
 & $i$ [deg] & \multicolumn{3}{c}{\textbf{48}$_{-2}^{+6}$}\\
 \hline
 Comptonization & $A_\text{compTT}$ [10$^{-3}$] & 4.01$_{-1.88}^{+0.90}$ & 5.17$_{-0.95}^{+0.01}$ & 3.65$_{-1.65}^{+18626.63}$ \\
 & $kT$ [keV] & 20.52$_{-9.53}^{+49.09}$ & 26.08$_{-1.77}^{+1.40}$ & 35.28$_{-33.28}^{+11.29}$ \\
 & $\tau$ & 0.52$_{-0.22}^{+0.37}$ & 0.63$_{-0.36}^{+1.07}$ & 0.01$_{-0}^{+0.15}$\\ 
 \hline
plasma & $A_\text{phot}$ & \multicolumn{3}{c}{\textbf{8.58}$_{-8.56}^{+41.89}$}
\\
 &$\log\xi$ & \multicolumn{3}{c}{\textbf{6.33}$_{-1.68}^{+0.67}$} \\ 
 \hline
cross calibration & $c_\text{XISf}$ & 1.00 & 1.00 & \nodata \\
 &$c_\text{XIS1}$ & 1.00 & 0.99$_{-0.01}^{+0.01}$ & \nodata \\
& $c_\text{XISf-PIN}$ & 1.16 & 1.26$_{-0.06}^{+0.07}$ &\nodata \\
\hline
individual & $\chi^2$/dof & 2232.56/[2106-11] & 1985.06/[2060-13] & 1128.64/[1100-11] \\
contributions & p-value & 0.02 & 0.83 & 0.20 \\\hline \hline
 & $\chi^2$/dof & \multicolumn{3}{c}{5346.275/[5276-34]} \\
 & p-value &  \multicolumn{3}{c}{0.16} \\
 & $\chi^2_\text{red}$ & \multicolumn{3}{c}{1.02} \\\hline\hline 
\end{tabular}
\end{center}
\end{table}

\begin{table}[!ht]
\begin{center}
\caption{Spectral Parameters for Fairall~9 multi-epoch fit with additional photoionized emission and two ionized reflection components; See the text for a description of the model. \textit{Suzaku} Spectra are normalized to XIS0 data. The power law normalization is $\text{photons}\,\text{keV}^{-1}\,\text{cm}^{-2}\,\text{s}^{-1}$ at 1\,keV.\label{mtable_double}}

\begin{tabular}{ccccc}
\hline 
& & Suzaku A & Suzaku B & XMM B \\
 \hline \hline
contiuum \& & $A_\text{pex}$ [$10^{-3}$] & 8.02$_{-0.38}^{+0.31}$ & 9.84$_{-1.66}^{+0.37}$ & 4.81$_{-1.23}^{+0.02}$ \\
cold reflection & $\Gamma$ & 2.07$_{-0.01}^{+0.01}$ & 2.17$_{-0.01}^{+0.01}$ & 2.09$_{-0.04}^{+0.04}$ \\
 & $R$ & 1.33$_{-0.19}^{+0.21}$ & 0.95$_{-0.16}^{+0.23}$ & 1.89$_{-0.16}^{+0.11}$ \\
 \hline
ionized reflection 1 &$A_\text{reflionx}$ [$10^{-4}$] & 0.51$_{-0.12}^{+0.00}$ & 0.41$_{-0.28}^{+0.36}$ & 0$_{-0}^{+0.02}$
\\
 & $Fe/Solar$ & \multicolumn{3}{c}{\textbf{0.75}$_{-0.07}^{+0.10}$} \\
& $\xi$ [erg\,cm\,s$^{-1}$] & 627.7$_{-411.3}^{+924.7}$ & 1091.9$_{-685.2}^{+5908.1}$ & 100$_{-0}^{+6900}$ \\
\hline
relativistic blurring 1 & $q$ & 10.0$_{-1.0}^{+0.0}$ & 10.0$_{-2.3}^{+0.0}$ & 10.0$_{-8.0}^{+0.0}$ \\
  &$a$ & \multicolumn{3}{c}{\textbf{0.97}$_{-0.01}^{+0.02}$}\\
 & $i$ [deg] & \multicolumn{3}{c}{\textbf{36}$_{-3}^{+3}$}\\
 \hline
ionized reflection 2 &$A_\text{reflionx}$ [$10^{-4}$] & 1.00$_{-0.35}^{+0.32}$ & 2.35$_{-0.43}^{+0.59}$ & 1.52$_{-1.02}^{+0.39}$\\
& $\xi$ [erg\,cm\,s$^{-1}$] & 12.1$_{-5.9}^{+8.6}$ & 1$_{-0}^{+1.2}$ &1$_{-0}^{+30.6}$ \\
\hline
relativistic blurring 2 & $q$ & 2$_{-0}^{+0.2}$ & 2.1$_{-0.1}^{+0.3}$ &5.6$_{-1.1}^{+4.4}$ \\
 \hline
plasma & $A_\text{phot}$ & \multicolumn{3}{c}{\textbf{0.09}$_{-0.03}^{+0.36}$}
\\
 &$\log\xi$ & \multicolumn{3}{c}{\textbf{4.3}$_{-0.0}^{+0.7}$} \\ 
 \hline
cross calibration & $c_\text{XISf}$ & 1.00 & 1.00 & \nodata \\
 &$c_\text{XIS1}$ & 1.00 & 0.99$_{-0.01}^{+0.01}$ & \nodata \\
& $c_\text{XISf-PIN}$ & 1.16 & 1.43$_{-0.08}^{+0.07}$ &\nodata \\
\hline
individual & $\chi^2$/dof & 2208.44/[2106-11] & 2012.66/[2060-13] & 1168.47/[1100-11] \\
contributions  & p-value & 0.04 & 0.67 & 0.05\\\hline \hline  
 & $\chi^2$/dof & \multicolumn{3}{c}{5389.6/[5276-34]} \\
 & p-value &  \multicolumn{3}{c}{0.08} \\
 & $\chi^2_\text{red}$ & \multicolumn{3}{c}{1.02} \\\hline\hline
\end{tabular}
\end{center}
\end{table}

\end{document}